\documentclass[reprint,
showpacs,preprintnumbers,twocolumn,
amsmath,amssymb,
 aps,prl,a4paper
]{revtex4}

\usepackage{graphicx}
\usepackage{dcolumn}
\usepackage{bm}
\usepackage{units}
\usepackage{color}

\DeclareGraphicsExtensions{.pdf,.png,.jpg}

\def\vec#1{\bm{#1}}
\newcommand{\revision}[1]{#1}
 
\begin{document}

\title{A mechanical model for guided motion of mammalian cells}

\author{Patrick Bitter}
\author{Kristof L. Beck}
\author{Peter Lenz}%
 \email{peter.lenz@physik.uni-marburg.de}
\affiliation{%
 Department of Physics, Philipps-Universit\"at Marburg and LOEWE Center for Synthetic Microbiology, Marburg, Germany 
}%

\date{\today}

\begin{abstract}
  We introduce a generic, purely mechanical model for environment-sensitive motion of mammalian cells that is applicable to chemotaxis,
  haptotaxis, and durotaxis as modes of motility. It is able to
  theoretically explain all relevant experimental observations, in
  particular, the high efficiency of motion, the behavior on
  inhomogeneous substrates, and the fixation of the lagging pole
  during motion. Furthermore, our model predicts that efficiency of
  motion in following a gradient depends on cell geometry (with
  more elongated cells being more efficient).
\end{abstract}

\maketitle

Motility and directed cell motion play an important role in many
biological processes ranging from embryonic development
\cite{Chicurel2002,Gray2003,Huttenlocher2011} to tissue invasion by
pathogenic microorganisms
\cite{Chicurel2002,Gray2003,Prost2008,Parsons2010} and cancer
progression
\cite{Lo2000,Gerisch2008,Bordeleau2010,Parsons2010}.

Often extracellular cues are used to regulate the decision in which
direction the cell will move
\cite{Macnab1972,Tindall2008,Levine2008}. Depending on these
cues one distinguishes between: (1) chemotaxis where
directed motion is guided by solvent chemical cues \cite{Macnab1972};
(2) haptotaxis where substrate-bound cues influence the cell-substrate
adhesiveness \cite{Carter1965}; and (3) durotaxis where mechanical
cues such as substrate rigidity influence the directed motion
\cite{Lo2000}.

Physics-based experimental and theoretical approaches to study these
phenomena have attracted considerable interest in the last
decade. Many studies have been devoted to bacterial chemotaxis
\cite{Tindall2008,Tindall2008a,
  Adler1966,Macnab1972,Adler1975,Tu2008,Zhu2012} and to
\textit{Dictiostelium discoideum} as model system for amoeboid
migration \cite{Fischer1989,Fuller2010,
  Hecht2011,Levine2008,Buenemann2010,Camley2014}. The main challenge is to
analyze and theoretically model the interplay between molecular
processes and the emerging macroscopic motion. For amoeboid motion
this is further complicated as shape changes have to be taken into
account \cite{Buenemann2010,Hecht2011,Camley2014}.

In contrast, the mesenchymal migration of mammalian cells has so far
not been studied theoretically. Although many details have been
characterized experimentally \cite{Zaman2005,Hacker2012,Gerisch2008}
theoretical studies have focused only on the cell shape during
migration \cite{Rubinstein2005}, on continuum descriptions for cell
populations \cite{Gerisch2008,Hacker2012} or on special short-term
aspects of migration like migration speed \cite{DiMilla1991} or
effective adhesiveness \cite{Zaman2005}.

\begin{figure}[tbp]
 \centering
 \includegraphics[width=\columnwidth]{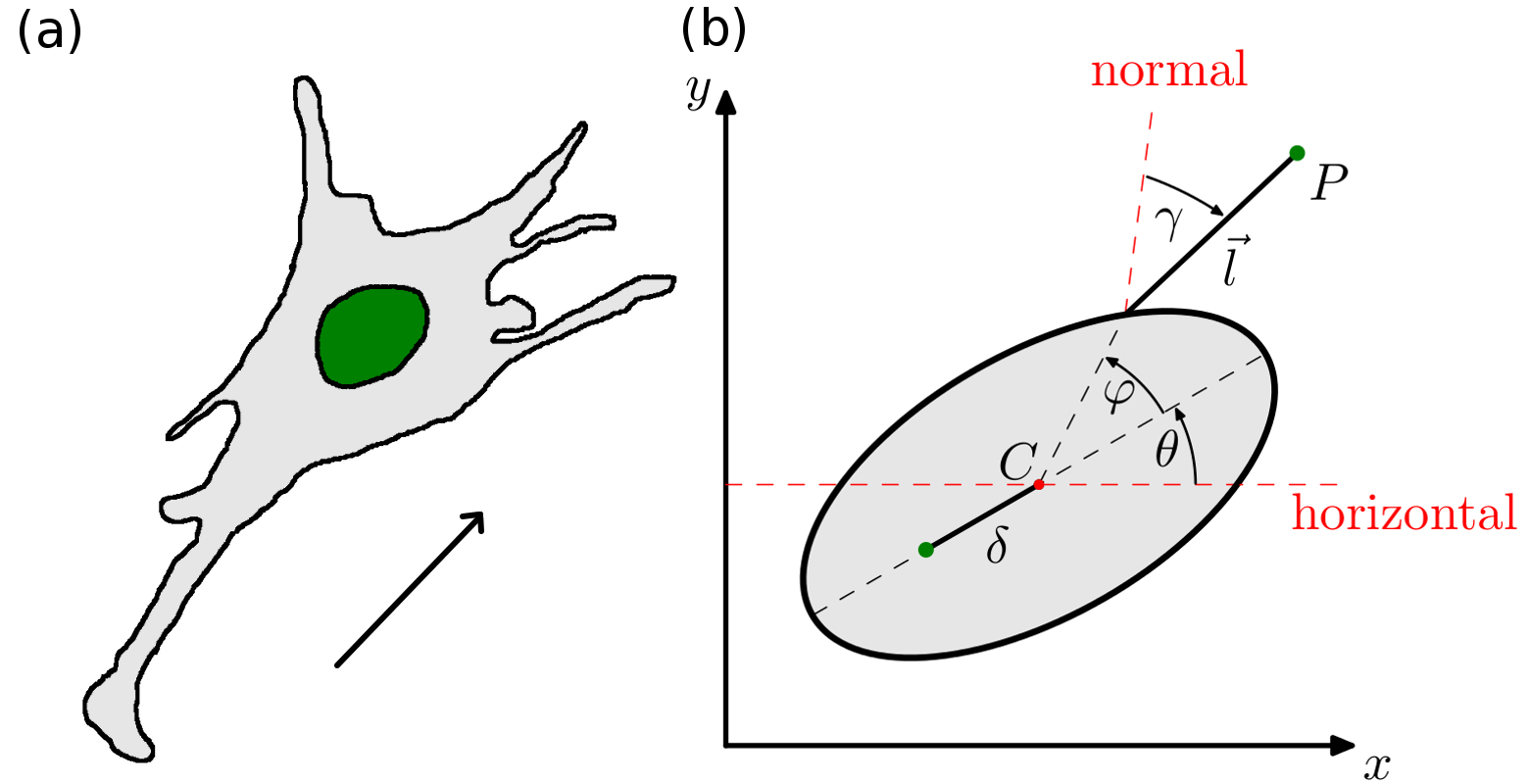}
 \caption{(a) Typical cell shape as observed in
   \cite{Lo2000} for a cell polarized in the direction of the
   arrow. The protrusions mainly grow into the forward
   direction. (b) In our model the polarized cell is
   represented by an ellipse, the attachments by lines. The position
   of an attachment point is defined by the length of a protrusion and
   three angles: the rotation angle of the cell $\theta$, the angle
   $\varphi$ for the attachment position on the cell membrane,
   and the angle $\gamma$ between the central line and the arm. The
   distance between the center of the ellipse and the \revision{anchor point} is $\delta$. \label{fig:scheme}}
\end{figure}

In this paper we study the motility of mammalian epithelial-like
cells. They are influenced by a variety of chemical and physical
signals, in particular by different mechanical forces
\cite{Butcher2009} and \revision{can show a very high efficiency
  in following all kinds of gradients \cite{Theveneau2010}.}
Mammalian cells predominantly migrate by a crawling motion
\cite{Keren2008}. It consists of a cycle of five discrete steps
carried out within about ten minutes
\cite{Mogilner2009,Lauffenburger1996,Pathak2011,Ridley2003,DiMilla1991,Zaman2005,Parsons2010,Huttenlocher2011,Chicurel2002,Keren2008}:
(1) polarization of the cell yielding a defined leading and a defined
lagging pole; (2) formation of protrusions \revision{(predominantly)}
at the leading pole and attachment of the lagging pole\revision{, see
  Fig.~\ref{fig:scheme}(a)}; (3) stabilization of these protrusions by
adhesion to the substrate or the extracellular matrix (ECM); (4)
translocation of the cell-body by myosin-mediated contraction; (5)
retraction of the rear by loosening the adhesions at the lagging
pole. This crawling is propelled by the active lamellipodeum at the
leading edge which pulls the passive cell body forward
\revision{\cite{Friedl2004}.}  Cell motility depends on the stiffness
of the substrate and the ECM. These effects are mediated indirectly
via their impact on cell shape \cite{Holmes2012} and directly through
so far unresolved mechanisms referred to as durotaxis
\cite{Lo2000,Trichet2012,Provenzano2008,Yeung2005,Gray2003}.

The mechanical interaction with the substrate or the ECM can be
thought of as a bilinear sequential binding \cite{Pathak2011}, which
affects the cell predominantly on the nanoscale through
mechanosensing mechanisms \cite{Butcher2009,Trichet2012}. However,
cells can also chemically manipulate the ECM, e.g., in case of cancer,
where tumor cells stiffen the surrounding ECM
\cite{Butcher2009,Provenzano2008}, and build a rigid stroma around the
tumor. This step in ECM stiffness then promotes cells from the outside
moving inside, but prevents cells from the inside to migrate outside
\cite{Bordeleau2013}. In general, mammalian cells are not passive
recipients of mechanical forces, \revision{but actively respond 
by pulling or pushing the ECM
\cite{Lauffenburger1996,Lo2000,Butcher2009}.} 

We introduce here a simple, generic model for environment-sensitive
motion of fibroblast-like cells. It is solely based on mechanics and
is applicable to chemotaxis, haptotaxis, and durotaxis as modes of
motility. It provides the first theoretical explanation of the high
efficiencies of mesenchymal-like motion independent of cellular
morphology. Our model also covers the motility dynamics on large time
scales. In particular, we can capture the statistical properties at
environmental discontinuities, e.g., a step in substrate stiffness or
a step in concentration of a chemo-attractant. These results indicate
that regulation of taxis might be based on mechanical forces.

In our model we represent the polarized cell body by an ellipse with
major radius $R_l$ and minor radius $R_s$. The orientation with
respect to the $x$-axis is measured by the angle $\theta$ that
parameterizes rotation around the center $C$. To counterbalance the
forces of the protrusions \revision{the cell is attached to the
  surface at an anchor point. Moving cells typically have a shape similar to
  the one shown in fig.~\ref{fig:scheme}(a) with a single elongated
  tail that appears when the cell-body moves forward.  This indicates
  that there is only a single anchor point. The motion occurs in such a way that the
  cell is effectively rotated around this anchor point}, see
Fig.~\ref{fig:scheme}(b). \revision{In principle, cells
  could also fix the position of the tail with several anchor
  points. Then, the cell is effectively rotated around these fixed
  points. For simplicity, we restrict here the analysis to a single
  anchor point. We assume that the anchor point
  remains at a fixed position while the cell is not moving. Real
  cells might change their shape during motion that could lead to a
  change in the position of the anchor point. However, we do not take
  such effects into account.}

In the following we assume that a molecular mechanism initially
polarizes the cell along the $x$-axis (for example in the presence of
a gradient as discussed below). For our purpose we do not explicitly
model this process and assume that it leads to a normally distributed
initial angle $\theta\left(t=0\right)$ with mean $\mu_\theta = 0$. The
standard deviation $\sigma_{\theta}$ was estimated by fitting a normal
distribution to the polarization model shown in \cite{Levine2013}.

The protrusions by which the cells pull themselves forward are
represented by adhesive arms that grow out of the ellipse at a random
angle $\varphi$ with probability distribution\revision{s given by Gaussian
distributions centered around $\varphi=0$ (for the leading pole) and
$\varphi=\pi$ (for the lagging pole)}
\begin{equation}
  \label{eq:pside}
  p(\varphi)
  =\frac{p_+}{\sqrt{2\pi}\sigma_+}e^{-\varphi^2/2
      \sigma_+^2}+
\frac{p_-}{\sqrt{2\pi}\sigma_-}e^{-(\varphi-\pi)^2/2
      \sigma_-^2}.
\end{equation}
The weights $p_+$ of the leading pole and $p_-=1-p_+\revision{< p_+}$ of the lagging
pole reflect the initial polarization of the cell and shift the arm
distribution towards the leading pole. 

Length and direction of the arms are random. For simplicity, we draw
the angle $\gamma$ between growth direction and surface normal from a
Gaussian distribution with mean $\mu_\gamma = 0$ and standard
deviation $\sigma_\gamma$. Similarly, the arm length $l$ is
distributed normally with mean $\mu_l$ and standard deviation
$\sigma_l$.

Arm formation occurs at constant rate. Every arm applies a linear
force on the cell that is proportional to the concentration of a
chemoattractant or rigidity of the substrate. Thus, if at time $t$
there are $N$ arms that are attached to the cell body at position
$\vec{x}_i=\vec{x}_i(\varphi_i)$ pointing in direction $\vec{l}_i$
(with $1 \leq i \leq N$) the total force on the cell is given by
\begin{equation}
 \vec{F}\left(\vec{r}_\mathrm{cell},\theta,t\right) 
 = \sum_{i=1}^N \vec{F}_i 
 = \sum_{i=1}^N k\left(c\right)\vec{l}_i\left(\vec{r}_\mathrm{cell},\theta,\varphi_i\right),
\end{equation}
where $k\left(c\right) = k_0 c\left(\vec{r}_\mathrm{arm}\right)$
depends on the chemoattractant concentration or the substrate rigidity
$c\left(\vec{r}\right)$. This results in a translocation of the
cell. Furthermore, the arms exert a total torque $ M\left(\vec{r}_\mathrm{cell},\theta,t\right) 
 = \sum_{i=1}^N \left( \vec{x_i} \times \vec{F_i} \right)_z$,
that leads to a rotation of the cell that can be interpreted as a
gradual repolarization of the cell.

For most of our simulations we used a linear gradient of fixed
strength $c_0 = {1}/{R_l}$. This results in a standard deviation
of the initial angle of $\sigma_{\theta} \approx 1$.

\begin{figure}[tbp]
 \centering
 \includegraphics[width=\columnwidth]{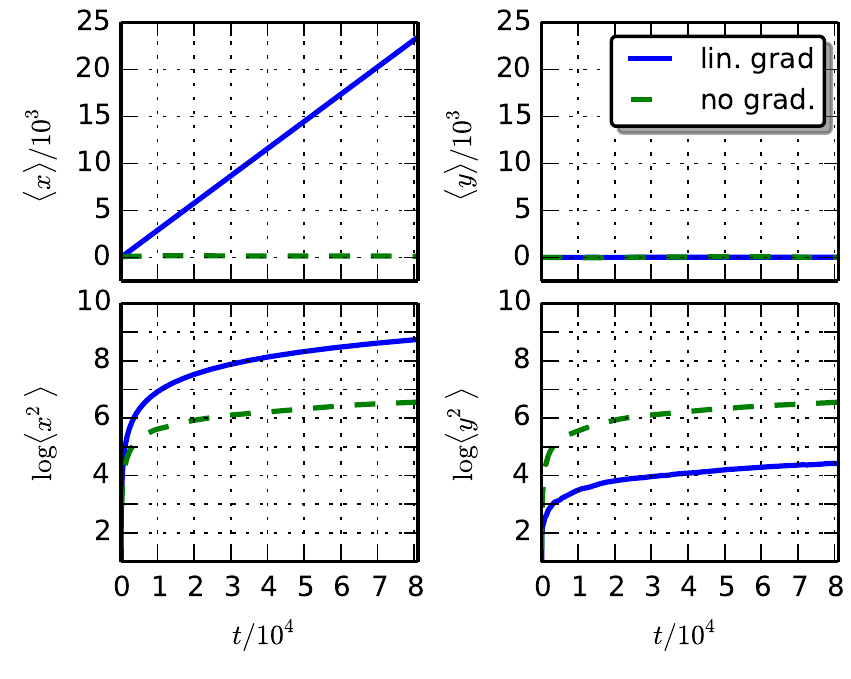}
 \caption{\revision{Classification of cell motion as obtained from the
     numerically calculated moments $\left\langle x \right\rangle$,
$\left\langle y \right\rangle$,$\left\langle x^2 \right\rangle$, and
$\left\langle y^2 \right\rangle$. In the absence of gradients the cell
performs a random walk characterized by vanishing $\left\langle x
\right\rangle$ and $\left\langle y \right\rangle$ and standard
deviations that increase with time. In the presence of a gradient this
random walk becomes biased. In this case the motion in $y$-direction
is suppressed compared with the case of no gradient. Data were obtained
by averaging over 500 independent runs. Length and time are given in
units of $R_l$ and $\delta t$, respectively.}
}
 \label{fig:moments}
\end{figure}

To classify the planar cell motion (in the $x$-$y$ plane), we
numerically calculated the moments $\left\langle x \right\rangle$,
$\left\langle y \right\rangle$,$\left\langle x^2 \right\rangle$, and
$\left\langle y^2 \right\rangle$, where $\langle ...\rangle$ denotes
the average over 500 independent runs, see Fig.~\ref{fig:moments}.  In these simulations
\revision{we implemented the above mentioned cycle of independent
  steps (1)-(5). Starting from the polarized shape [step (1)]} we grow
$N$ new arms of lengths $l_i$ at angles $\varphi_i$ in every iteration
\revision{[steps (2) and (3)]}. The new position and orientation of the
cell is then obtained by solving
$\vec{F}\left(\vec{r}_\mathrm{cell},\theta,t\right) = 0$ and
$M\left(\vec{r}_\mathrm{cell},\theta,t\right)=0$ independently
\revision{[step (4)]}. Then, all arms are removed \revision{[step
  (5)]}, the time is increased \revision {by $\delta t$} and the
iteration starts over.  In the absence of a gradient the cells perform an
isotropic random walk. For $N = 10$ (for parameter values see
\cite{Supp}) the effective diffusion coefficients $D_x \approx D_y =
43.77{R_l^2}/{\delta t}$ were measured by fitting linear functions to
$\langle x^2 \rangle$ and $\langle y^2 \rangle$.

In this limit the model can also be solved analytically. From the
one-dimensional probability densities of $\varphi$, $\gamma$ and $l$
the two-dimensional probability density of the force induced by an arm
can be calculated \cite{Supp}.  
We have compared these analytical
results with those obtained by direct numerical integration of the
model and found excellent agreement. It is interesting to note that
this force probability density resembles the shapes of migrating
lamellipodial domains of keratocytes \cite{Rubinstein2005}. If one
assumes that these shape deformations reflect the forces then the
forces acting on the cell in our simple probabilistic model are
remarkably similar to those exerted by the actin cytoskeleton of
keratocytes.
 
Next, we consider substrates with gradients. For a linear gradient 
$c_0 = {1}/{R_l}$ parallel to the $x$-axis, the cells perform a biased
random walk in the gradient direction while the motion in the perpendicular
direction is suppressed $\sim 135$-fold ($D_y = 0.32{R_l^2}/{\delta
  t}$ in presence of a gradient compared to $D_y =
43.77{R_l^2}/{\delta_t}$ in absence of a gradient)\revision{, see Fig.~\ref{fig:moments}.}
To quantify the efficiency of the motion in following the applied gradient we measured
the chemotactic factor
\begin{equation}
 CF = \left\langle \frac{L_\mathrm{grad}}{L_\mathrm{tot}} \right\rangle.
\end{equation}
Here, $L_\mathrm{tot}$ is the total path length and $L_\mathrm{grad}$
the length of the projection in the direction of the gradient.

To investigate the robustness of our model to varying gradients $c_0$
and its behavior for small gradients we looked at the dependence of
$CF$ on $c_0$\revision{, see Fig.~\ref{fig:gains} in \cite{Supp}.} 
For small gradients we see a strong increase in
efficiency with the gradient strength, but the efficiency saturates fast
to its maximum value of around $CF = 95 \%$. 

\begin{figure}[tbp]
 \centering
 \includegraphics[width=\columnwidth]{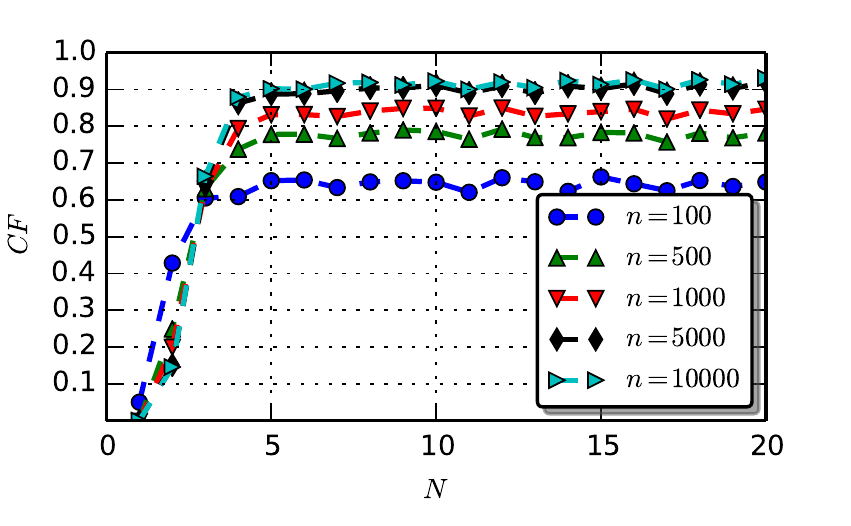}
 \caption{\revision{Efficiency of motion (as quantified by the chemotactic
   factor $CF$) as function of the number of arms $N$ for different simulation
   time steps $n$.}
}
 \label{fig:arms}
\end{figure}

An increase in number of arms results in a speedup of motion and
an increase in efficiency, see Fig.~\ref{fig:arms}. However, there is saturation in efficiency
and speed for large numbers of arms. If we take into account that 5 to
10 arms with an average length of $7 \mu \textup{m}$ would roughly
cover between 10 and 20\% of the surface of the cell (each arm has
about $22 \mu \textup{m}^2$ surface area while the whole cell body has
about $1000 \mu \textup{m}$ membrane area
\cite{Raucher1999,Baxter2002}) we reach a good balance between
increase of membrane area and gain of efficiency within this range of
$N$. This number is also comparable to the number of arms found
experimentally \cite{Lo2000}.

\begin{figure}[tbp]
 \centering
 \includegraphics[width=\columnwidth]{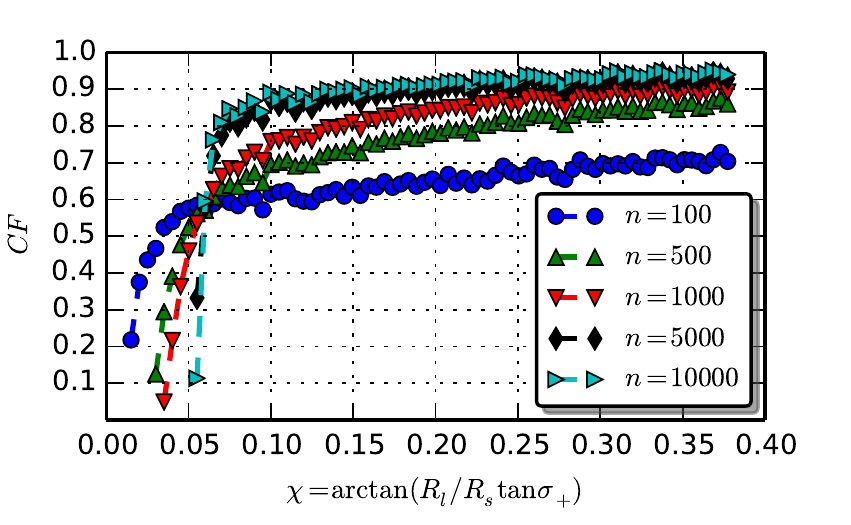}
 \caption{\revision{Chemotactic
   factor $CF$} is shown as function of cell geometry characterized by
   $R_l/R_s$. More elongated cells follow with higher efficiency the
   gradient (of strength $c_0=1/R_l$). The data shown are for different
   simulation time steps $n$. \revision{For smaller $n$ (i.e. earlier
     simulation times), $CF$ still depends on the initial alignment of
     the cells. With increasing $n$ the cells align with the gradient
     increasing in this way $CF$. However, as can be seen only cells
     with $\chi > 0.06$ (i.e. cells which are sufficiently elongated) are
     able to align with the gradient.}
}
 \label{fig:eff_wofit}
\end{figure}

Next, we analyzed the influence of the cell geometry on the efficiency
of motion. As Fig.~\ref{fig:eff_wofit} shows, $CF$ depends on the
geometry of the cell characterized by the ratio $R_l/R_s$. Thus, more
elongated cells (with $R_l>R_s$) have a higher $CF$. This is somewhat
surprising as these cells have a broader force distribution than less
elongated cells. However, as we show in \cite{Supp} $CF$ depends predominantly on the ability of the cell to align with the prescribed
gradient. Thus, for more elongated cells this higher ability
compensates for the broader force distribution. The $CF$ as
determined by the ability to align with the gradient (characterized by
a rotational rate $\alpha$) is given by \cite{Supp}
\begin{equation}
	CF = \exp(x^2)
  	\left[ \textup{erf}\left (\pi/(\sqrt 2 \sigma_\theta)-x\right )
	+ 	\textup{erf}(x) \right].
\end{equation}
Here, $x=2^{-1/2}\sigma_\theta \alpha/\chi$ and the opening angle of
the force distribution $\chi=\arctan\left(R_l/R_s\tan
  \sigma_+\right)$ parameterizes the dependence on cell
geometry.

Lo et.\ al. \cite{Lo2000} have shown that non-moving cells grow longer
protrusions on stiff substrates than they do on soft substrates. This
implies a regulative effect of the substrate rigidity. To account for
this effect in our model we assign each arm $i$ the length
\begin{equation}
l_i+ \delta l_i\left(c\right) = l_i+ \frac{\delta l_\mathrm{max}}{1+\frac{k_c}{c}},
 \label{eq:elongation}
\end{equation}
where the regulated elongation $\delta l_i\left(c\right)$ depends on
the stiffness (or concentration or adhesiveness) $c =
c\left(\vec{r}_\mathrm{arm}\right)$ at the position of the arm
$\vec{r}_\mathrm{arm}$.

If we increase the average arm length, we see an increase in
efficiency that saturates for longer arms\revision{, see Fig.~\ref{fig:armreg} in \cite{Supp}.} This rise in efficiency comes with an increase in
speed. If we use the concentration-dependent regulation of arm lengths
we see a 1\% increase in maximum efficiency compared to the
unregulated system, but the same efficiency is reached with an average
arm length up to 40\% shorter compared to the unregulated system.

The above results are robust with respect to variations in the
standard deviations of the distributions for $\varphi$ and
$\gamma$. $CF$ remains nearly constant for $\sigma_+$ and
$\sigma_\gamma$ in a range from $\sim 0.01\pi$ to $0.2\pi$. For even
broader distributions we see a decrease in efficiency as a result of
insufficient polarization of the cell. Experimentally, it has been
observed that the protrusions grow almost in the normal direction out of
the surface of the cells, and that the protrusions are located around
a narrow region at the leading pole
\cite{Maly2001,Svitkina2003,Mogilner2005,Parsons2010}. We find a
similar behavior (and the associated high efficiencies in motion) only
for distributions with small standard deviations indicating that
$\varphi$ and $\gamma$ are tightly regulated.

\begin{figure}
 \centering
 \includegraphics[width=\columnwidth]{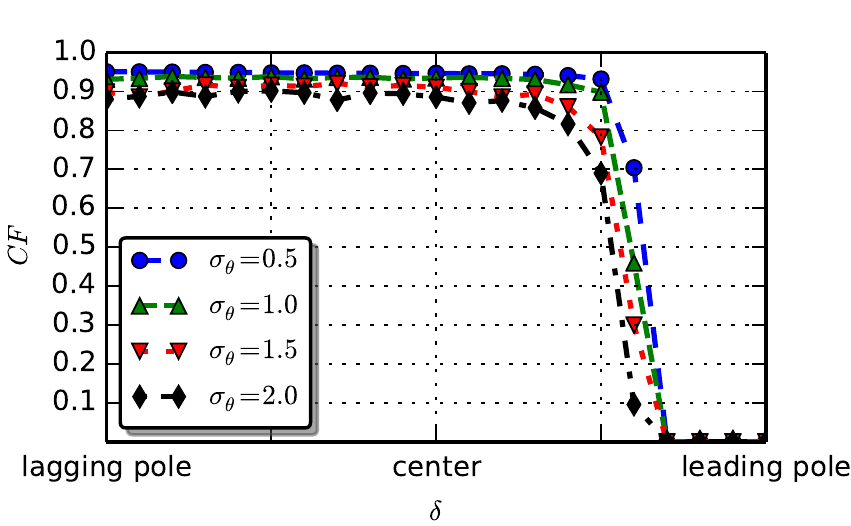}
 \caption[Dependency on the shift $\delta$.]{Dependence of $CF$ on the
   position of the \revision{anchor} point to the substrate. The distance of
   this point from the center of the ellipse is denoted by
   $\delta$. As this point moves from the center of the ellipse
   towards the leading pole, $CF$ drops sharply to zero. This
   indicates that in order to be able to follow a prescribed gradient
   the cells need to have a defined leading pole and an \revision{anchor}
   point sufficiently far away from this pole.\label{fig:delta}}
\end{figure}

Furthermore, cell motion shows an interesting dependence on the position
of the \revision{anchor} point of the cell to the substrate quantified by the
parameter $\delta$. If this point is shifted towards the leading pole,
\revision{i.e. closer to the protrusions at the front, the torque
  exerted by these arms is reduced due to the shorter lever arm and}
the efficiency drops sharply to zero indicating that the cell is not
able to follow the gradient at all (see Fig.~\ref{fig:delta}). \revision{This
indicates that the efficiency in following a gradient is dominated by
protrusions close to the leading pole.}

Cells often encounter inhomogeneous substrates. As a general scenario,
we analyze the movement towards a step in substrate rigidity where
crawling cells show an interesting behavior. At these steps (that
could represent the transition from a rigid stroma of a tumor to the
softer surroundings \cite{Butcher2009}) cells tend to move from the
softer substrate to the stiffer substrate \cite{Bordeleau2013}.
\begin{figure}[tbp]
\centering
 \includegraphics[trim=1.6cm 1.6cm 2cm 1.3cm,clip=true, width=\columnwidth]{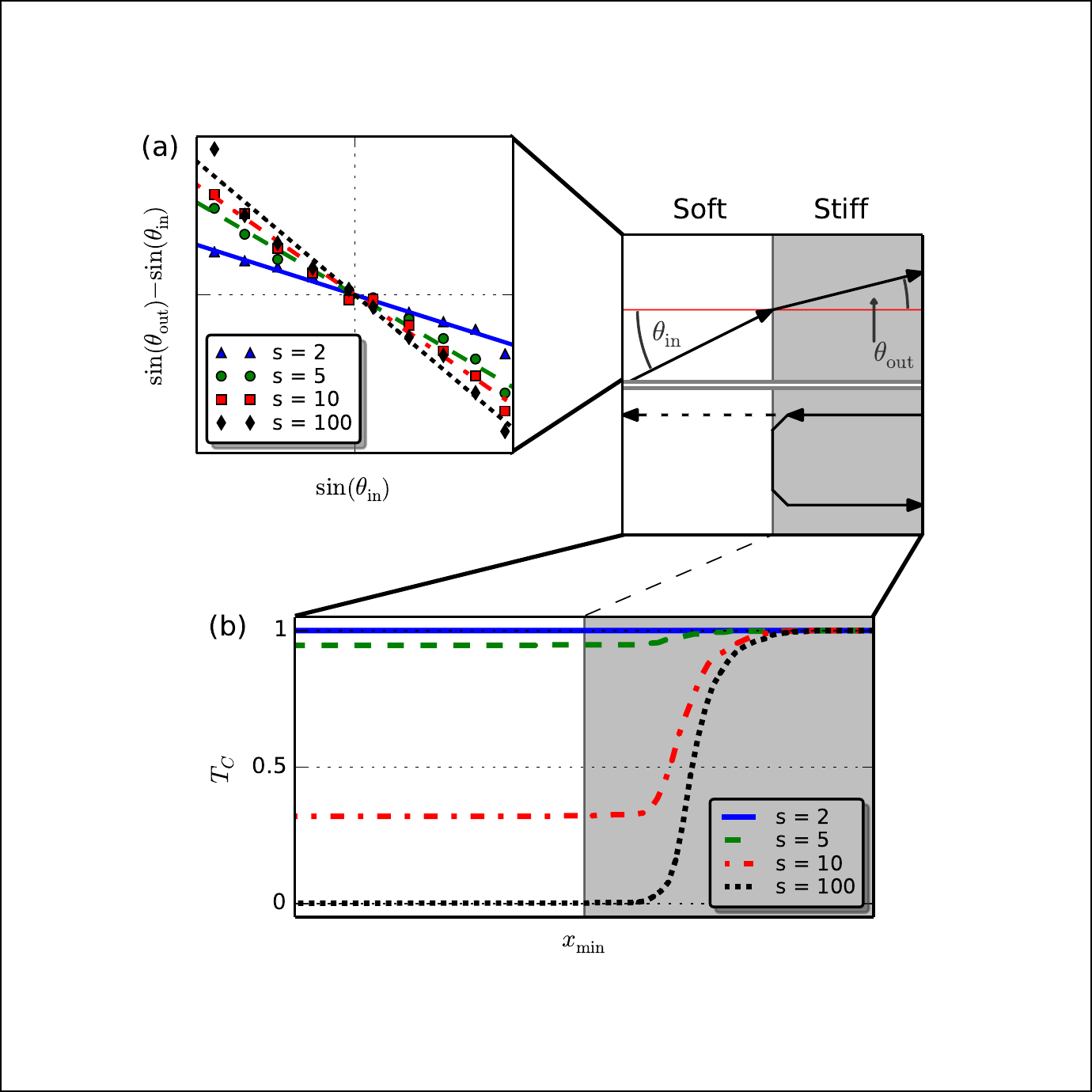}
 \caption[Transition properties at a sharp concentration/stiffness
 step.]{Transition at a sharp step. (a)
   If cells move from the soft region to the stiff region one finds a
   refractive behavior depending on the step size $s$. (b) Cells moving from the stiff to the
   soft region experience a barrier that they overcome with a step-size-dependent transmission coefficient. For large step sizes no
   cells cross the barrier while for small step sizes
   almost all cells are able to cross it. 
\label{fig:step}}
\end{figure}
Cells moving in this direction are only weakly influenced by the step.
They keep moving but their trajectory bends towards the direction
perpendicular to the step, see Fig.~\ref{fig:step}. We observed that
the relation between the angle of the cell before and after the step
obeys a refraction law similar to that of light allowing us to
characterize the motion by refraction indices, see
Fig.~\ref{fig:step}(a). The ratio of refraction indices
${n_\mathrm{Soft}}/{n_\mathrm{Stiff}}=\sin \theta_{\mathrm out}/\sin
\theta_{\mathrm in}$ decreases with increasing
step size $s=c_1/c_0$, where $c_0$ and $c_1$ are the stiffness
of the softer and stiffer region, respectively.

On the other hand, a step from a stiff substrate to a softer substrate
represents a barrier for cells. The passing probability depends on the
step height. From the distribution of the minimal $x$-positions
encountered by the cells during 500 iterations, we can calculate the
probability of a cell moving across the step, which we define as
transmission coefficient $T_\mathrm{C} =
\int_{-\infty}^{x_\mathrm{step}} p\left(x\right) =
p\left(x<x_\mathrm{step}\right)$, where $p\left(x\right)$ is the
probability to find a cell that traveled to position $x$, and
$x_\mathrm{step}$ is the position of the step, see
Fig.~\ref{fig:step}(b). This coefficient decreases as the step size
increases showing that the barrier effect becomes much stronger for
larger steps\revision{, see Fig.~\ref{fig:intrusion} in \cite{Supp}.}

\begin{figure}[tbp]
\centering
\includegraphics{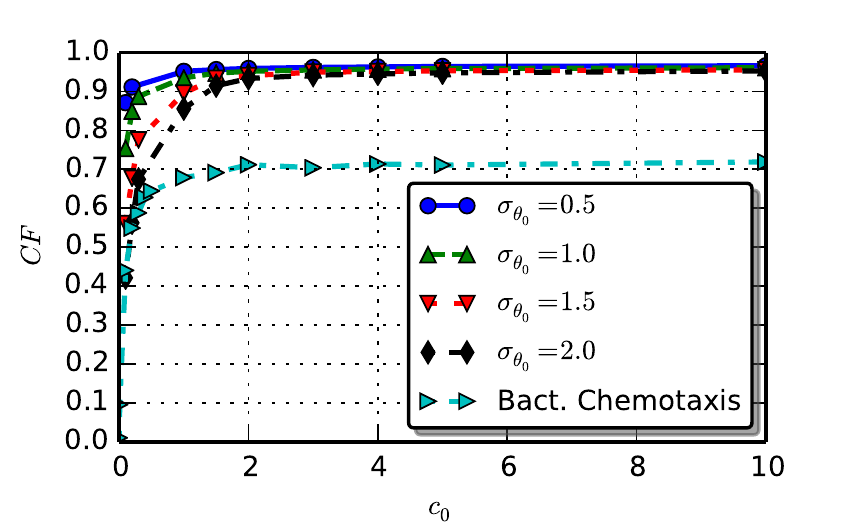}
\caption{Variation of the efficiency with the gradient strength
  $c_0$. The cells reach their maximum efficiency at rather small
  gradients compared to a swim-tumble model of bacterial chemotaxis
  (based on \cite{Macnab1972,Tindall2008}).\label{fig:comp}}
\end{figure}

Finally, \revision{we wanted to compare our results with other types
of cellular motion. However, there are neither theoretical nor
experimental data available for the efficiency of mammalian cell
motion as function of gradient strength. We therefore decided to
compare our model with data for swim-tumble chemotaxis, the standard
model for cellular motion. More specifically we tested our results
against} a two-dimensional swim-tumble model for bacterial
chemotaxis with a parameter set optimized for efficiency
\cite{Macnab1972,Tindall2008,Supp}.  As Fig.~\ref{fig:comp} shows our
model yields significantly higher efficiencies than the swim-tumble
model.

\revision{To summarize, in this study we have introduced a generic mechanical
model for environment-sensitive motion of mammalian cells. Motion
occurs by polarized growth of protrusions which push and rotate the
cell. The description of molecular interactions occurs on a coarse-grained
level effectively entering into the probability distribution for arm
growth, see Eq.~(\ref{eq:pside}). It is not the goal of this simplified model to achieve a
detailed (molecular) description of cellular motility. Rather, we
introduce it to analyze
the influence of mechanical forces on the regulation of fibroblast
motion.}

\revision{The agreement of our findings with the experimental observations
indicates that mechanical forces indeed play a significant role in
this process. More specifically, the observed high efficiency in
following a gradient is a robust feature of our model (see
figs.~\ref{fig:eff_wofit} and \ref{fig:comp}). We find for a large parameter range chemotactic factors
close to 1 as observed experimentally in
 \cite{Theveneau2010}. Furthermore, the results on the motion in
inhomogeneous environments are in good agreement with the experimental
observations at steps in substrate rigidity \cite{Bordeleau2013}.}

\revision{Our analysis identifies two geometrical factors that have a
significant impact on efficiency of cell motion: the position of the
anchor point and the geometry of cell quantified by the ratio of major
to minor radius. For both quantities we make specific theoretical
predictions (figs.~\ref{fig:eff_wofit} and \ref{fig:delta}) that are
experimentally testable.}

\revision{
Furthermore, the moments $\langle x\rangle$, $\langle x^2\rangle$,
$\langle y\rangle$ and $\langle y^2\rangle$ can be easily measured for
individual cells of different geometry for different gradients. By
comparing these data with our theoretical predictions (Fig.~\ref{fig:comp} and
Figs.~\ref{fig:armreg} and \ref{fig:class_armreg} in \cite{Supp}) information can be obtained about the concentration dependent regulation of arm lengths. } To check our results concerning steps in
concentration, adhesiveness and stiffness, one could use the methods
presented in \cite{Bordeleau2013} to produce flat substrates of
different stiffness and measure the polarization axes of the cells
before and after the interface as well as the transmission
coefficients with time-lapse microscopy.

\revision{There are many possible extensions of our model. In future
  work we will take into account} the mechanical effects that the
cells have on the substrate. If cells attach protrusions to the
substrate and contract, they locally stiffen the substrate. This local
stiffening of the substrate might lead to an effective attraction
between two cells in proximity, in this way promoting aggregation.



\newpage

\renewcommand\thesection{S}
\renewcommand\thetable{\thesection\arabic{table}}
\onecolumngrid
\setcounter{figure}{0}
\setcounter{table}{0}
\setcounter{equation}{0}
\renewcommand\thefigure{\thesection\arabic{figure}}
\renewcommand\theequation{\thesection.\arabic{equation}}

\begin{center}
\textbf{\large Supplemental Material: A mechanical model for guided motion of mammalian cells}
\end{center}

\section{Analytical results}

\subsection{Force distribution}
\label{sec:force-distribution}

\begin{figure}[hbt]
 \centering
 \includegraphics[width=.5\textwidth, trim=2cm 22.5cm 10cm 2cm,
 clip]{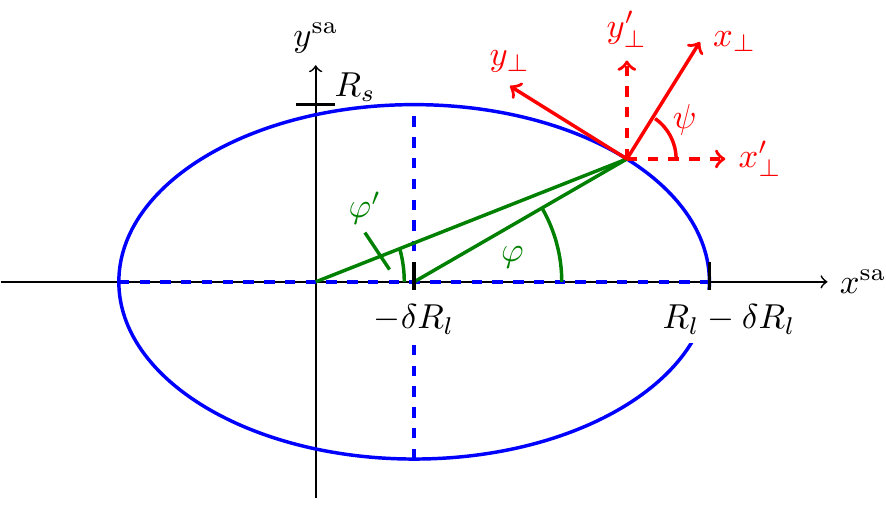}
 \caption{Ellipse in the coordinate system with fixed counterpart of
   the protrusions as origin and the semimajor axis as abscissa.
   Vectors to the attachment point from the origin and from the center
   of the ellipse are shown in green. The coordinate systems in the
   attachment point are shown in red: solid lines for the system with
   one axis perpendicular to the surface, dashed lines for the system
   rotated parallel to the system of the semiaxis.}
 \label{fig:elli_sketch}
\end{figure}

The cell is represented by an ellipse with semiaxis $R_l \ge R_s$ (see Fig.~\ref{fig:elli_sketch}). An adhesive arm grows out of the ellipse at angle $\varphi$.
We now choose a coordinate system $\left(x^\textup{sa}, y^\textup{sa}\right)$ with the fixed
counterpart of the protrusions as origin and the semimajor axis as abscissa.
The vector to the origin of an arbitrary arm is given by
\begin{equation}
 	\vec r_\textup{arm} = 
		\left(
			\begin{array}{c}
				R_l ~(\cos \varphi - \delta) \\
				R_s \sin \varphi
			\end{array}
		\right).
\end{equation}

The angle between the vector to the attachment point and the abscissa is
\begin{equation}
	\varphi^{\prime} = \arctan\left(\frac{R_s \sin(\varphi)}{R_l
	~(\cos\varphi - \delta)}\right).
\end{equation}

Length $l$ and direction $\gamma$ of the arms are Gaussian distributed
\begin{align}
	p(\gamma)
	= \frac{1}{\sqrt{2 \pi} \sigma_\gamma} 
	\exp\left[-\frac 1 2 \left( \frac{\gamma}{\sigma_\gamma} \right)^2 \right]
	, \quad 
	p(l)
	= \frac{1}{\sqrt{2 \pi} \sigma_l} 
	\exp\left[-\frac 1 2 \left( \frac{l -
	\overline l}{\sigma_l} \right)^2 \right].
\end{align}
Here, the angle $\gamma$ is measured relative to the normal of the ellipse in
the attachment point. The probability to find an arm with length $l$ and
angle $\gamma$ is
\begin{align}
	p(l, \gamma)
	= p(l) p(\gamma)
	= \frac{1}{\sqrt{2 \pi} \sigma_\gamma} 
	\exp\left[-\frac 1 2 \left( \frac{\gamma}{\sigma_\gamma} \right)^2 \right]
	\frac{1}{\sqrt{2 \pi} \sigma_l} 
	\exp\left[-\frac 1 2 \left( \frac{l -
	\overline l}{\sigma_l} \right)^2 \right].
\end{align}

The force caused by an arm with length $l$ and angle $\gamma$ is given by 
\begin{equation}
	\vec F^\perp = k l
	\left(
		\begin{array}{c}
			\cos \gamma \\
			\sin \gamma
		\end{array}
	\right)
	\equiv
	\left(
		\begin{array}{c}
			F^\perp_x \\
			F^\perp_y
		\end{array}
	\right),
\end{equation}
in the coordinate system $(x_\perp, y_\perp)$ in the attachment point (see Fig.~\ref{fig:elli_sketch}). In the absence of a gradient we can assume $k = 1$.

The probability to have a force $(F^\perp_x, F^\perp_y)$ is
\begin{align}
	p^\perp(F^\perp_x, F^\perp_y) 
	=
	\frac{1}{2 \pi \sigma_l \sigma_\gamma}
	\frac{1}{\sqrt{{F^\perp_x}^2 + {F^\perp_y}^2}}
	\exp\left[-\frac 1 2 \left( \frac{\sqrt{{F^\perp_x}^2 + {F^\perp_y}^2} -
	\overline l}{\sigma_l} \right)^2 \right]
	\exp\left[-\frac 1 2 \left( \frac{\arctan(F^\perp_y,
	F^\perp_x)}{\sigma_\gamma} \right)^2 \right],
\end{align}
where we denote by $\arctan(u,v)$ the arc tangent of $\frac u v$ taking into
account the quadrant of the point $(u, v)$.

To get the force in the coordinate system of the semiaxes, we need to rotate by
the angle 
\begin{equation}
	\psi = \arctan\left(R_l \sin\varphi, R_s \cos\varphi\right).
\end{equation}
Thus,
\begin{align}
	p_\varphi^\textup{sa}(F_x, F_y)
	&= p_\varphi^\perp(F^\perp_x, F^\perp_y)
	\equiv p^\perp(F_x ~\cos\psi + F_y ~\sin\psi, -F_x ~\sin\psi +
        F_y ~\cos\psi) \nonumber
	\\ \nonumber
	&=
	\frac{1}{2 \pi \sigma_l \sigma_\gamma} 
	\frac{1}{\sqrt{F_x^2 + F_y^2}}
	\exp\left[-\frac 1 2 \left( \frac{\sqrt{F_x^2 + F_y^2} -
	\overline l}{\sigma_l} \right)^2 \right]
  	\\ 
  	&\times
	\exp\left[-\frac 1 2 \left( \frac{\arctan(-F_x ~R_l ~\sin\varphi + F_y ~R_s ~\cos\varphi,
	F_x ~R_s ~\cos\varphi + F_y ~R_l ~\sin\varphi)}{\sigma_\gamma} \right)^2 \right].
	\label{eq:si_pFphi}
\end{align}

\begin{figure}[tbp]
 \centering
 \includegraphics{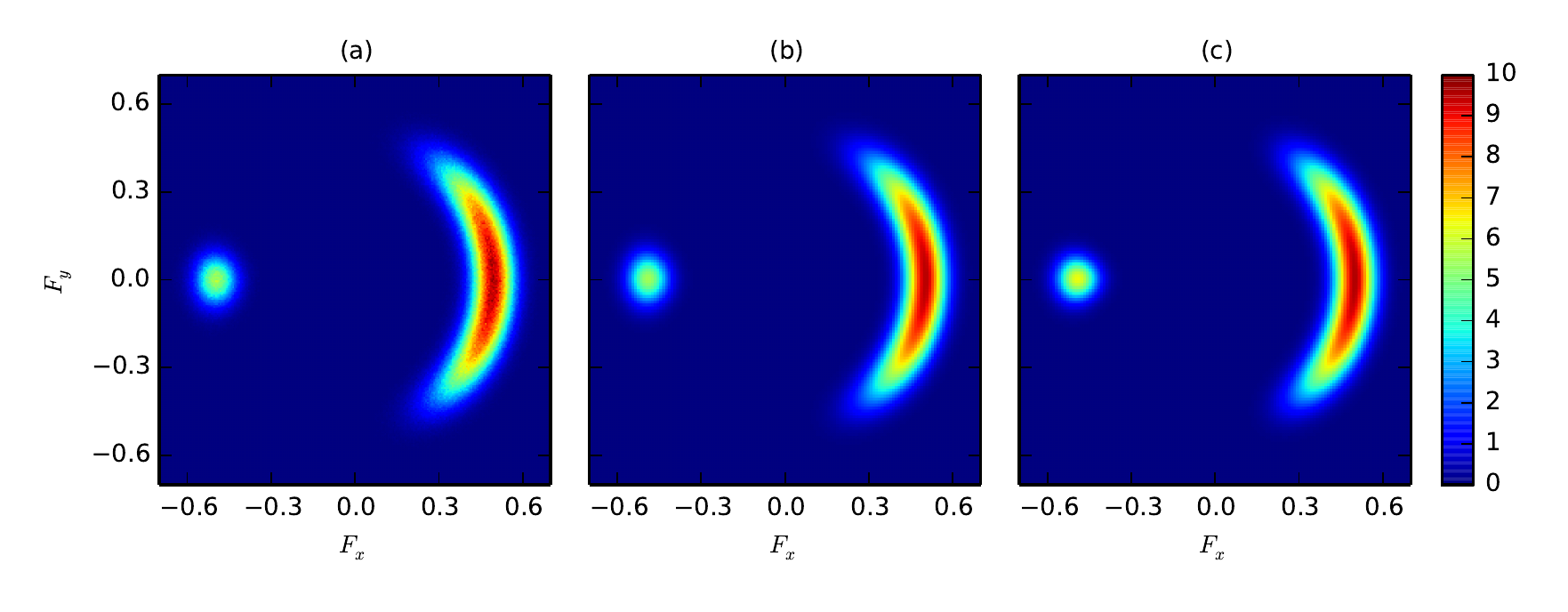}
 \caption{Probability density for the force $(F_x, F_y)$. (a)
   Numerical result; (b) analytical result \eqref{eq:psa} (for
   same set of parameters); (c) analytical result in the approximation \eqref{eq:psa_approx}.}
 \label{fig:psa_FxFy_ana_num}
\end{figure}

This probability density is valid for any (fixed) $\varphi$. The complete
density is given by
\begin{align}
	p^\textup{sa}(F_x, F_y)
	&= \int\limits_{-\pi}^{\pi} d\varphi ~p_\varphi^\textup{sa}(F_x, F_y) ~p(\varphi),
	\label{eq:psa}
\end{align}
which has to be calculated numerically. The result shown in Fig.~\ref{fig:psa_FxFy_ana_num} is in good agreement with the numerical data. 

As Fig.~\ref{fig:psa_FxFy_ana_num} shows the distributions of $\gamma$
and $\varphi$ are rather sharp (i.e. $\sigma_{\varphi-} <
\sigma_\gamma \ll 1$). Therefore, we can approximate the corresponding
Gaussian functions by delta functions leading to 
\begin{align}
	\nonumber
	p^\textup{sa}(F_x, F_y)
	\nonumber
	&\approx
	\frac{p_+}{2 \pi \sigma_l \sigma_{\varphi+}}
	\frac{R_l ~R_s ~\sqrt{F_x^2 + F_y^2}}{R_l^2 F_x^2 + R_s^2 F_y^2}
	\exp\left[-\frac 1 2 \left( \frac{\sqrt{F_x^2 + F_y^2} -
	\overline l}{\sigma_l} \right)^2 \right]
	\exp\left[-\frac 1 2 \left( \frac{\arctan\left( R_s ~F_y, R_l
	~F_x\right)}{\sigma_{\varphi+}} \right)^2 \right]
	\\ \label{eq:psa_approx}
	&+ 
	\frac{p_-}{2 \pi \sigma_l \sigma_\gamma} 
	\frac{1}{\sqrt{F_x^2 + F_y^2}}
	\exp\left[-\frac 1 2 \left( \frac{\sqrt{F_x^2 + F_y^2} -
	\overline l}{\sigma_l} \right)^2 \right]
	\exp\left[-\frac 1 2 \left( \frac{\arctan\left(-F_y, -F_x
	\right)}{\sigma_\gamma} \right)^2 \right].
\end{align}

From this approximation, we conclude that the width of the distribution is given
by $\sigma_l$ in radial direction and by $\sigma_{\varphi+}$ and
$\sigma_\gamma$ in angular direction. We define
$\chi$ as the angle, for which $\arctan\left( R_s ~F_y, R_l
  ~F_x\right) \leq \sigma_{\varphi+}$, i.e.
\begin{align}
	\chi = \arctan \left( \frac{R_l}{R_s}
 	~\tan\sigma_{\varphi+} \right).
\end{align}
Thus, $\chi$ quantifies the ability of the cell
to move straight ahead and/or turn towards a given gradient. It depends on
$\sigma_{\varphi+}$ and on the geometric factor $\frac{R_l}{R_s}$. 

\subsection{Model for motion along gradients}
\label{sec:openingangle_follow}
To investigate the influence of the opening angle on the ability to follow a gradient we analyze a somewhat simplified scenario: we neglect all effects from the back
of the cell and do not take into account the torque. 

In a first step we describe the motion as a random walk. In doing so we keep track
of the position and direction of cell. In each step the cell follows its direction for a
constant distance and then changes direction by an angle $\pm \eta$ relative to
its current direction.
The probabilities for selecting $+\eta$ and $-\eta$ are denoted by $p$ and $1 -
p$, respectively. The probability that after $N$ steps the cell is rotated $n$ times by $\eta$ and $N-n$ times by $-\eta$ (resulting in a net rotation of $(2n-N) \eta$) is given by
\begin{align}
	P_N(n) &= \binom{N}{n} ~p^n ~(1 - p)^{N-n}.
\end{align}
For large numbers ($N
\gg 1$, $n \gg 1$, $N - n \gg 1$) this becomes a continuous Gaussian
distribution for the net rotation $x = (2n-N) \eta$
\begin{align}
	P(x) &= \frac{1}{\sqrt{2 \pi} \sigma} 
	\exp\left[ -\frac 1 2 \frac{\left(x - \overline x\right)^2}{\sigma^2} \right],
\end{align}
with mean $\overline x  = N (2 p - 1) \eta$ and standard deviation $\sigma = 2 \sqrt{N p (1 -p)} \eta$.

\begin{figure}[tbp]
 \centering
 \includegraphics{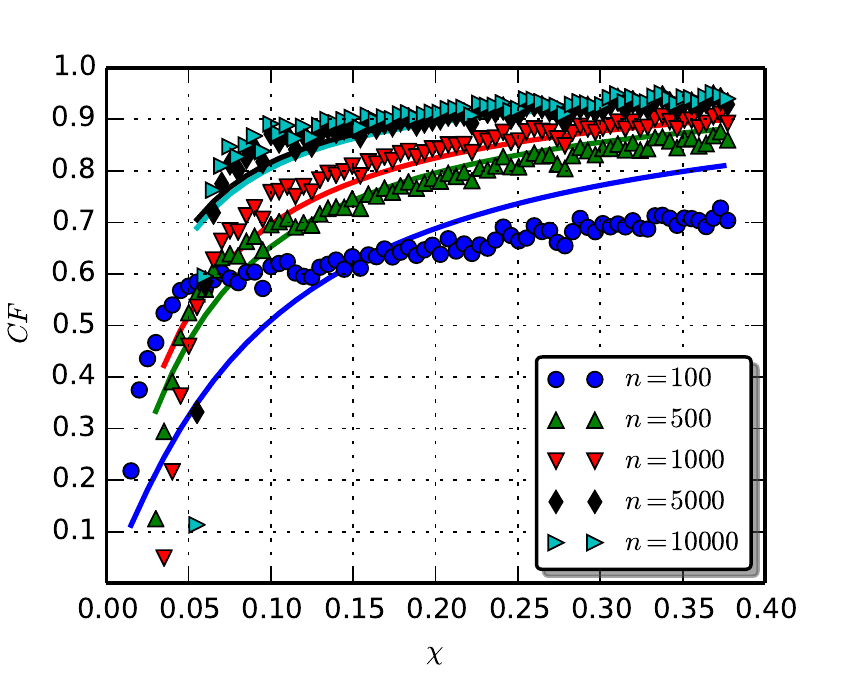}
 \caption{Efficiency of motion as function of $\chi$ and number of time steps $n$ in the simulation. 
At early simulation stages the efficiency is dominated by the alignment with
 the 
gradient. At later times all cells are
 aligned with the gradient and the efficiency approaches 1.
For vertically elongated ellipses ($R_s > 2 R_l$) the torque can turn the
 cell against the gradient resulting in a negative efficiency. If this
 event occurred the simulation was stopped. 
The lines show the results of fitting Eq.~\eqref{eq:eff_fit} to the
 numerical data.}
 \label{fig:eff_fit}
\end{figure}

\begin{figure}[hbt]
 \centering
 \includegraphics[width=0.48\textwidth]{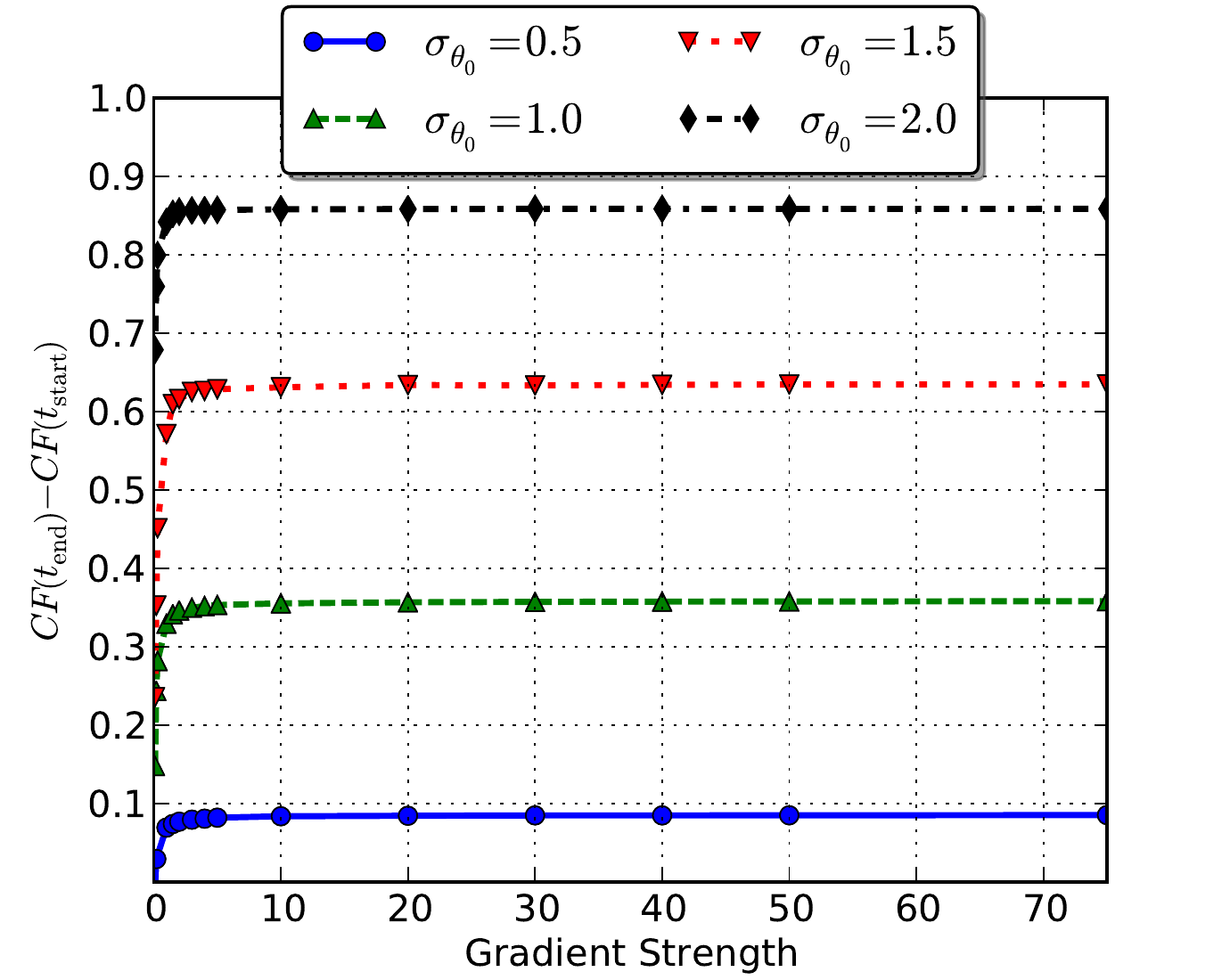}
 \caption{
  Gains in chemotactic factor $CF$, i.e. difference between $CF$ at
  the end and at the beginning of the simulation, as function of
  gradient strength $c_0$. Results are shown for different standard
  deviations $\sigma_\theta$ of the initial polarization angle
  $\theta_0$. For smaller $\sigma_\theta$ the cells are initially
  aligned more in the direction of the gradient, so less gain is
  expected.
   \label{fig:gains}}
\end{figure}

\begin{figure}[hbt]
 \centering
 \includegraphics[width=0.48\textwidth]{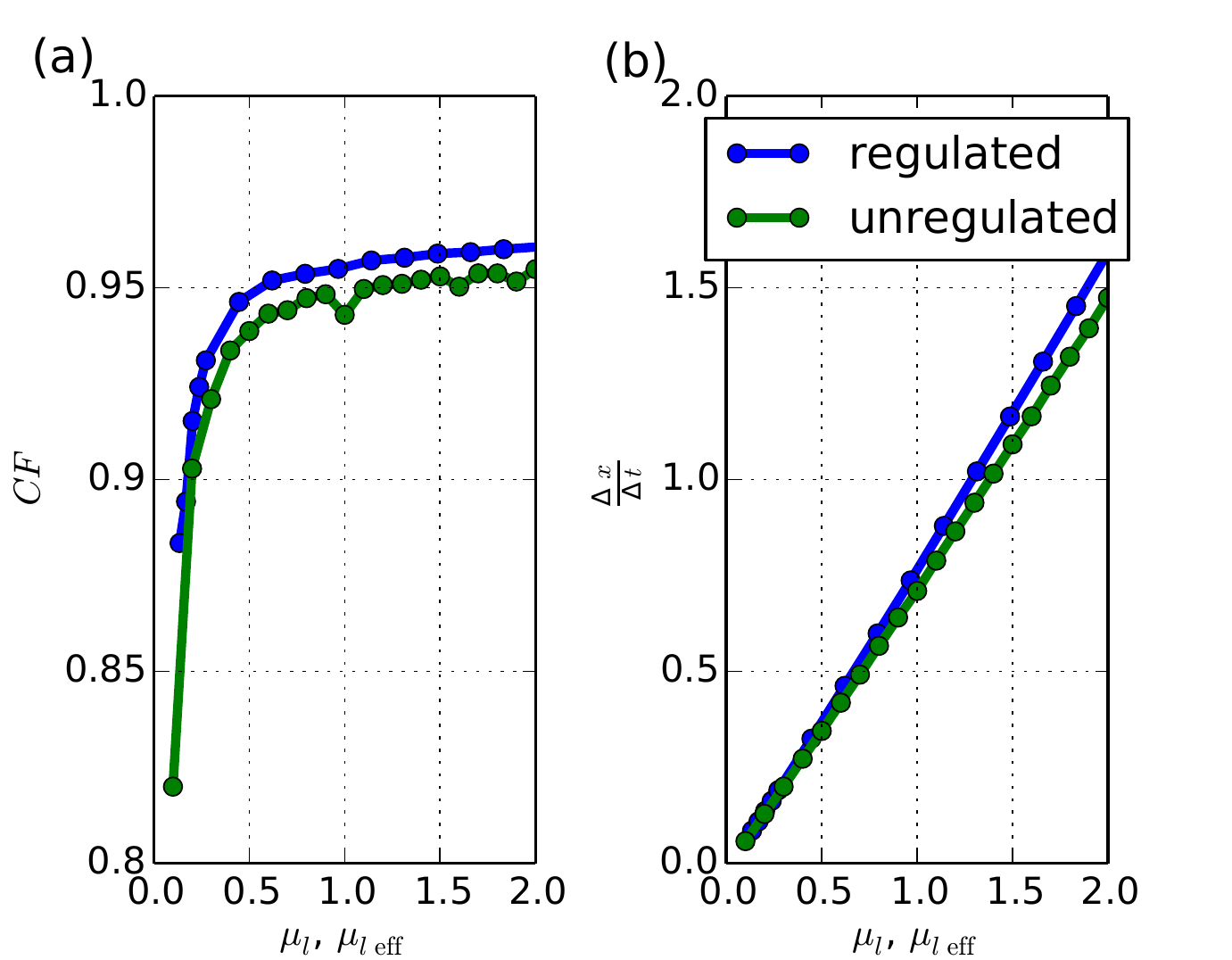}
  \caption{
  Effect of the concentration dependent regulation of arm
  lengths on efficiency of motion. (a) If we increase the average
  arm length $\mu_l$ we see an increase in efficiency that saturates
  for longer arms. If we use the concentration dependent regulation
  of arm lengths ($\mu_{l,\textup{eff}}$, see
  Eq. (\ref{eq:elongation})) we get a 1\% increase in maximum
  efficiency compared to the unregulated system from $CF \approx
  0.95$ to $CF \approx 0.96$, but the same efficiency is reached
  with an average arm length up to 40\% shorter compared to the
  unregulated system.  
  (b) Speed of motion, characterized by the covered path
  length $\Delta x$ per time unit $\Delta t$ increases with increasing
  arm length. Additionally, for identical average arm length motion is
  faster with regulation. The difference in speed increases with 
  increasing average arm length.
   \label{fig:armreg}}
\end{figure}

\begin{figure}[hbt]
 \centering
 \includegraphics[width=0.48\textwidth]{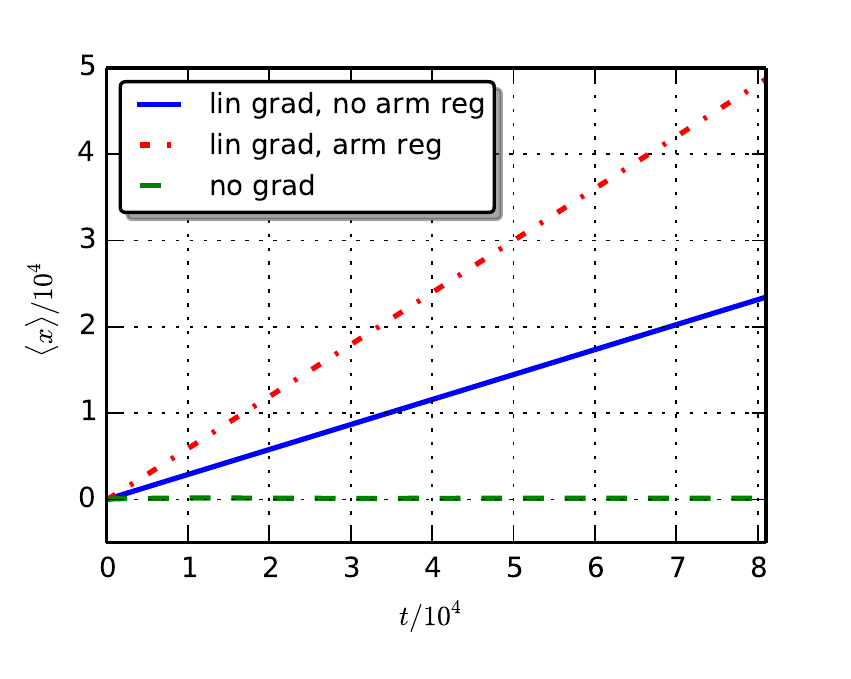}
 \caption{
 Classification of the migration dynamics. The cell performs
   a random walk which is biased in the presence of a gradient, see
   Fig.~\ref{fig:moments}. If the arms are regulated the motion in
   absence of a gradient stays the same, whereas the motion in
   presence of a gradient becomes faster at constant efficiency (additional red
   dash-dotted line). 
   \label{fig:class_armreg}}
\end{figure}

\begin{figure}[hbt]
 \centering
 \includegraphics[width=0.48\textwidth]{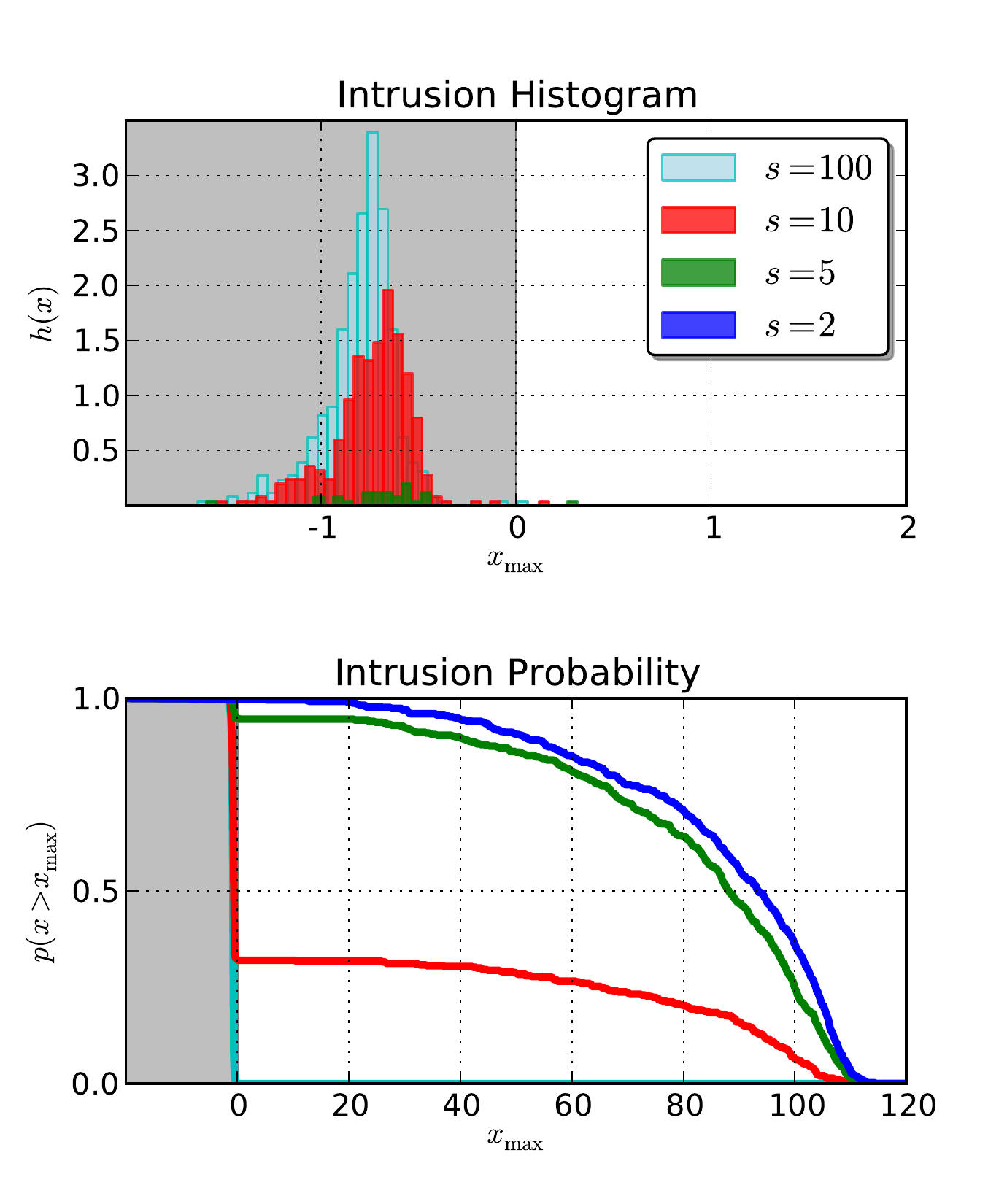}
  \caption{
  Histogram and probability of intrusion at a surface step
  (i.e. interface between  stiff and soft substrate region). Cells move from left
  (grey, stiff substrate) to right (white, soft substrate). The
  histogram shows the frequency of the maximal (most right) $x$
  value reached. We only show the region close to the step to focus
  on the cells that did not cross it.  
  Cells that have passed the step (most of the cells in simulations
  for small steps, i.e. blue and green) would be found at
  $x_\textup{max} > 2$, so they are not shown in the histogram.  
  For larger step sizes not all cells pass the step. The maximal distance
  traveled in positive $x$-direction peaks closer to the interface
  with decreasing step size. If the step size becomes too small almost
  all cells pass the step as indicated by the intrusion
  probability that is defined as the probability that a cell travels
  further than $x_\textup{max}$. 
   \label{fig:intrusion}}
\end{figure}

The efficiency of the motion can be quantified by the expectation value of $\cos x$ which
represents the fraction of the distance covered in the direction of a
gradient in $x$-direction 
\begin{align}
	\langle \cos x \rangle 
	= \int\limits_{-\infty}^\infty dx ~\cos x ~\frac{1}{\sqrt{2 \pi} \sigma} 
	\exp\left( -\frac 1 2 ~\frac{x^2}{\sigma^2} \right)
	= \exp\left(-\frac 1 2 \sigma ^2 \right).
\end{align} 
In the last equation one can identify the standard deviation $\sigma$ with the opening angle $\chi$ of the force distribution. Thus, this simple model predicts that with increasing $\chi$ the efficiency should decrease. As can be seen from Fig.~\ref{fig:eff_fit} this clearly contradicts our numerical findings. This indicates that the ability to follow the gradient is not determining the efficiency of motion. As we show now, it in fact depends crucially on the ability of the cell to align with the gradient.

To do so, we assume that initially the cell has an angle
$\theta_0$ between its major axis and the applied linear
gradient. With a probability $p$ the cell now rotates by
an angle $\Psi$ towards the gradient. Thus, after $n
= \frac{|\theta_0|}{\Psi}$ the cell is perfectly aligned with the gradient. 

Here, the efficiency can be quantified by the averaged probability of the cells to align
with the gradient
\begin{align}
	\int\limits_{-\pi}^\pi d\theta_0 ~p^{\frac{|\theta_0|}{\Psi}}
= 2 ~\frac{\Psi}{\ln p} ~\left[ \exp\left(\frac{\pi}{\Psi} ~\ln p\right) -
	1 \right].
\end{align}
Taking into account that the starting angles $\theta_0$ are Gaussian distributed
in our simulations, the efficiency becomes
\begin{align}
	\int\limits_{-\pi}^{\pi} d\theta_0 
	\frac{1}{\sqrt{2 \pi} \sigma_\theta} 
	~&\exp\left(-\frac 1 2 ~\frac{\theta_0^2}{\sigma_\theta^2} \right) 
	~p^{\frac{|\theta_0|}{\Psi}} \nonumber
 	= \frac{2}{\sqrt{2 \pi} \sigma_\theta} ~\int\limits_{0}^{\pi}
 	d\theta_0 ~\exp\left(-\frac 1 2 ~\frac{\theta_0^2}{\sigma_\theta^2} 
 	+ \frac{\theta_0}{\Psi} ~\ln p\right)
 	\\
 	\label{eq:eff_fit}
 	&=  
 	\exp\left[\frac{\sigma_\theta^2 ~(\ln p)^2}{ 2 ~\Psi^2} \right] 
 	~\left\{ \textup{erf}\left[\frac{1}{\sqrt 2 \sigma_\theta}
 	~\left(\pi 
 	- \frac{\sigma_\theta^2 ~\ln p}{\Psi} \right) 
 	\right]
 	+
 	~\textup{erf}\left[
 	~\left(\frac{\sigma_\theta ~\ln p}{\sqrt 2 ~\Psi} \right) 
 	\right]
 	\right\}.
\end{align}

Thus, the probability depends on the rotation
angle $\Psi$ (that we identify with the opening angle $\chi$ of the
force distribution) and on the probability $p$ (which depends on the strength of
the gradient). Because the exact relation between $p$ and the strength
of the gradient is not known we take 
$\ln p$ as a fitting parameter.
As one can see from Fig.~\ref{fig:eff_fit} the fitted curves
match the numerical data quite well for sufficiently large number of
simulation steps. At early simulation stages the numerical data
depends still on the initial orientation of the cell. 

For vertically elongated cell ($R_s > 2 R_l$) the gradient eventually
turns the cell into the ``wrong'' direction. This tilt in the wrong
direction increases with time since arms growing in this direction are
favored by the gradient. This process leads to an alignment of the
ellipse against the gradient. We stopped our simulations when such an
event occurred.

\section{Model for Bacterial Chemotaxis\label{sec:bac_model}}

To model the bacterial chemotaxis, we use the well-studied swim-tumble
model. The cell swims into a random direction for a normally
distributed length $l$. It measures the change in concentration $\Delta c$ of an attractant along this path. On the basis of $\Delta c$ the decision is made whether the cell maintains the current direction or tumbles. The former occurs with probability
\begin{equation}
 p\left(\Delta c\right) = \left\{ \begin{matrix} 
     \dfrac{p_0}{1-\dfrac{\Delta c (1-p_0)^2}{kp_0}} & \Delta c \leq 0
 \\ \dfrac{1}{1+\dfrac{k(1-p_0)}{\Delta c+p_0 (k- \Delta c)}}& \Delta c > 0\end{matrix} \right.
\end{equation}
which is larger than a basal probability of $p\left(\Delta c = 0\right)
= p_0$  for $\Delta c >0$ and smaller than $p_0$ for $\Delta c <
0$. The parameter $k$ quantifies the stiffness of the response. 

To compare this model with our model, we also applied here a concentration gradient with a linear slope in $x$-direction.

\clearpage
\section{Supplementary Figures}

\clearpage
\section{Supplementary Tables}

\begin{table}[hbt]
\begin{tabular}{cccccccccccc}
\hline \hline
Parameter 
& Fig.~\ref{fig:moments}~\&~\ref{fig:class_armreg}  
& Fig.~\ref{fig:arms} 
& Fig.~\ref{fig:eff_wofit}~\&~\ref{fig:eff_fit} 
& Fig.~\ref{fig:delta} 
& Fig.~\ref{fig:step}(a)
& Fig.~\ref{fig:step}(b) 
& Fig.~\ref{fig:comp} 
& Fig.~\ref{fig:psa_FxFy_ana_num}
& Fig.~\ref{fig:gains} 
& Fig.~\ref{fig:class_armreg}
& Fig.~\ref{fig:intrusion}
\\ \hline 
Iter. & 81000 & -- & -- & 500 & 500 & 500 & 500 & 1 & 500 & 500 & 500\\
Runs & 500 & 500 & 1000 & 500 & 500 & 500 & 500 & 5000000 & 500 & 500 & 500\\
$a$ & 1 & 1 & 1 & 1 & 1 & 1 & 1 & 1 & 1 & 1 & 1\\
$b$ & 0.5 & 0.5 &-- & 0.5 & 0.5 & 0.5 & 0.5 & 0.5 & 0.5 & 0.5 & 0.5\\
$\delta$ & -1 & -1 & -0.9 & -- & -1 & -1 & -1 & -0.9 & -1 & -1 & -1\\
$\sigma_\theta$ & 1.047 & 1.047 & 1.047 & -- & -- & 0 & 1.047 & -- & -- & 1.047 & 0\\
$k$ & 1 & 1 & 1 & 1 & 1 & 1 & 1 & 1 & 1 & 1 & 1\\
$c_0$ & -- & 1 & 1 & 1 & 1 & 1 & 1 & -- & -- & 1 & 1\\
$N$ & 10 & -- & 10 & 10 & 10 & 10 & 10 & 1 & 10 & 10 & 10\\
$p_+$ & 0.9 & 0.9 & 0.9 & 0.9 & 0.9 & 0.9 & 0.9 & 0.9 & 0.9 & 0.9 & 0.9\\
$\mu_l$ & 0.5 & 0.5 & 0.5 & 0.5 & 0.5 & 0.5 & 0.5 & 0.5 & 0.5 & -- & 0.5\\
$\sigma_l$ & 0.05 & 0.05 & 0.05 & 0.05 & 0.05 & 0.05 & 0.05 & 0.05 & 0.05 & 0.05 & 0.05\\
$\sigma_+$ & 0.5 & 0.1 & 0.1 & 0.5 & 0.5 & 0.5 & 0.5 & 0.3 & 0.5 & 0.5 & 0.5\\
$\sigma_-$ & 0.05 & 0.01 & 0.01  & 0.05 & 0.05 & 0.05 & 0.05 & 0.03 & 0.05 & 0.05 & 0.05\\
$\sigma_\gamma$ & 0.7 & 0.1 & 0.1 & 0.1 & 0.1 & 0.1 & 0.1 & 0.1 & 0.1 & 0.1 & 0.1\\
\hline \hline
\end{tabular}
\caption{Parameter values for the results shown in Figs.~2-7, S2-S7}
\end{table}

\begin{table}[htb]
 \begin{tabular}{ccc}
  \hline \hline
  Step size $s$ & $\frac{n_\mathrm{Soft}}{n_\mathrm{Stiff}}$ & $T_\mathrm{C}$ \\
  \hline
  2 & 0.9374 & 1.0000 \\
  5 & 0.8843 & 0.9480 \\
  10 & 0.8622 & 0.3220 \\
  100 & 0.8330 & 0.0020 \\
  \hline\hline
 \end{tabular}
 \caption{Values for the ratio of refraction indices $\frac{n_\mathrm{Soft}}{n_\mathrm{Stiff}}$ and transmission coefficient $T_\mathrm{C}$. The refractive effect becomes stronger with increasing step size whereas the transmission decreases.\label{tab:step_data1}}
\end{table}

\end{document}